\documentclass[twocolumn,prb,showpacs,color,superscriptaddress,epsfig,floatfix,amsmath,amssymb,amsfonts]{revtex4}
\usepackage{graphicx}
\usepackage{color}
\usepackage{multirow}

\begin{document}

\title{Coupled frustrated quantum spin-1/2 chains with orbital order in volborthite Cu$_3$V$_2$O$_7$(OH)$_2\cdot$2H$_2$O}

\author{O. Janson}
\email{janson@cpfs.mpg.de}
\affiliation{Max-Planck-Institut f\"{u}r Chemische Physik fester
Stoffe, D-01187 Dresden, Germany}

\author{J. Richter}
\affiliation{Institut f\"{u}r Theoretische Physik, Universit\"{a}t
Magdeburg, D-39016 Magdeburg, Germany}

\author{P. Sindzingre}
\affiliation{Laboratoire de Physique Th\'eorique de la
Mati\`ere Condens\'ee,\ Univ. P. \& M. Curie, Paris, France}

\author{H. Rosner}
\email{rosner@cpfs.mpg.de}
\affiliation{Max-Planck-Institut f\"{u}r Chemische Physik fester
Stoffe, D-01187 Dresden, Germany}

\date{\today}

\begin{abstract}
We present a microscopic magnetic model for the spin-liquid candidate
volborthite Cu$_3$V$_2$O$_7$(OH)$_2\cdot$2H$_2$O.
The essentials of this DFT-based model are (i)
the orbital ordering of Cu(1) $3d_{3z^2-r^2}$ and Cu(2)
$3d_{x^2-y^2}$ (ii) three relevant couplings
$J_{\mathsf{ic}}$, $J_1$ and $J_2$ (iii) the ferromagnetic
nature of $J_1$ and (iv) frustration governed by the
next-nearest-neighbor exchange interaction $J_2$.
Our model implies
magnetism of frustrated coupled chains in contrast to
the previously proposed anisotropic kagome model. Exact
diagonalization studies reveal agreement with experiments.
\end{abstract}


\maketitle

\section{Introduction}
The search for new magnetic ground states (GS) is a major
subject in solid state physics. Magnetic monopoles in the
spin ice system
Dy$_2$Ti$_2$O$_7$~(Ref.~\onlinecite{monopoles_theory, Dy2Ti2O7_monopoles_1, Dy2Ti2O7_monopoles_2}),
the metal-insulator transition in the spin-Peierls compound
TiOCl~\onlinecite{TiOCl_MIT_theory} and the quantum critical
behavior in
Li$_2$ZrCuO$_4$~\onlinecite{Li2ZrCuO4_chiT_CpT_DFT_simul,FHC_Li2ZrCuO4_DFT_simul} are
among recent discoveries that demonstrate the power of
combining precise experimental techniques with modern
theory. However, for a rather large number of problems
experiment and theory don't keep abreast, since it is often
tricky to find a real material realization for a well
studied theoretical model.  The most remarkable example is
the concept of a ``resonating valence bond''\cite{GS_RVB}
---  a magnetic GS formed by pairs of coupled spin-singlets
lacking the long range magnetic order (LRO). Subsequent
studies revealed a fascinating variety of disordered
GS,\cite{geom_fr_Greedan, geom_frustr_short_review}
commonly called ``spin liquids'' in order to emphasize their
dynamic nature, and even raised the discussion of their
possible applications.\cite{macro_entanglement}

\begin{figure}[tb]
\includegraphics[width=5.0cm]{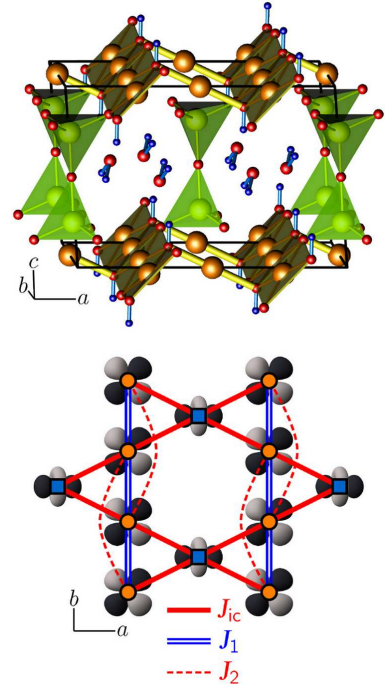}
\caption{\label{str}(Color online) Top: Cu(1)O$_2$
dumbbells (yellow), Cu(2)O$_4$ plaquettes (dark yellow),
V$_2$O$_7$ pyrovanadate groups (connected green tetrahedra) and H$_2$O molecules
in volborthite. Only short Cu--O bonds are
shown. Bottom: a distorted kagome layer in the crystal
structure of volborthite. The magnetically active orbitals
and leading exchange couplings are shown.}
\end{figure}

Following the common belief that the spin liquid GS may
emerge from the interplay of low-dimensionality, quantum
fluctuations and magnetic frustration, considerable effort
has been spent on the search for \mbox{spin-1/2} Heisenberg
magnets with kagome geometry. The synthesis of
herbertsmithite Cu$_3$Zn(OH)$_6$Cl$_2$~\onlinecite{herb}, the
first inorganic \mbox{spin-1/2} system with ideal kagome
geometry, and subsequent studies revealed besides the
desired absence of magnetic LRO\cite{herb_INS_chiT} (i)
intrinsic Cu/Zn structural disorder and (ii) the presence of
anisotropic interactions complicating the spin
physics.\cite{herb_ESR_DM}  The recently synthesized
kapellasite\cite{kapellasite_synth}
was predicted to imply modified kagome physics due to an
additional relevant coupling.\cite{kapel_hayd_PRL}

Since the search for a system representing the pure kagome model is far from
being completed, it is natural to consider systems with lower symmetry where
the distortion is small enough to keep the essential
physics.\cite{sqlat_anisotr_model}  This way, the attention has been drawn to
the mineral volborthite Cu$_3$V$_2$O$_7$(OH)$_2\cdot$2H$_2$O, where the Cu
sites form a slightly distorted kagome network.\cite{volb} However, the local
environment of two independent Cu sites is essentially different: Cu(1) forms
dumbbells of two short Cu--O bonds (and four long Cu--O bonds; ``2+4''), while
Cu(2) resides in a plaquette formed by four short bonds (Fig.~\ref{str}, top).
Recently, DFT studies of CuSb$_2$O$_6$, implying the ``2+4'' local environment
of Cu atoms, revealed that orbital ordering (OO) drastically changes the nature
of the magnetic coupling from three-dimensional (3D) to
one-dimensional(1D).\cite{CuSb2O6_DFT}  The crucial impact of OO onto magnetism
of volborthite will be in the focus of this paper.

The availability of a pure powder and negligible structural
disorder in volborthite inspired thorough experimental
studies of its magnetic properties.  The magnetic GS was
recently investigated by $^{51}$V~nuclear magnetic resonance
(NMR),\cite{volb_NMR_2} following earlier
NMR,\cite{volb_NMR_1} high-field electron spin
resonance\cite{volb_high_field_ESR} and muon spin
relaxation\cite{volb_muSR_chiT_CuZnx_2} studies.  The GS is
characterized by the absence of magnetic LRO, high density
of low energy excitations and two distinct scales of spin
fluctuations.  At 4.5 Tesla, volborthite undergoes a
transition to another magnetic phase. The fingerprint of
this transition is a step-like feature in the magnetization
curve. Similar features are observed at 25 and 46 Tesla,
hinting at a series of successive
transitions.\cite{volb_magn_high_field}  Between 60 and 70
Tesla, the slope of magnetization diminishes indicating a
possible onset of a magnetization plateau.\cite{footnote1}
Magnetic susceptibility measurements yield a broad maximum
at temperatures much smaller than the Curie-Weiss
temperature, indicating strong frustration.

Extensive experimental information stimulated theoretical
studies aiming to find a consistent description for
magnetism of volborthite. Since the pure kagome model
doesn't account for the experimental data, the studies were
focused on the GS and thermodynamical properties of the
anisotropic kagome model (AKM). However, attempts to reach
consistency by varying the degree of anisotropy were not
satisfying so far. The most striking disagreement is the
deviation of the theoretical magnetic susceptibility $\chi$
even at rather high temperatures ($T{\sim}J$).\cite{Sindzingre_07074264}

This disagreement originates from the choice of the AKM
which was based on geometry only, while the structural
peculiarities of volborthite were not considered.
A standard tool to treat such peculiarities properly
is density functional theory (DFT) calculations that can
provide a reliable microscopically based
model.\cite{TiOCl_MIT_theory, Li2ZrCuO4_chiT_CpT_DFT_simul,
kapel_hayd_PRL,HC_CuNCN_DFT,HC_CuSiO3H2O_DFT_QMC_chiT}  Here, we show that DFT calculations yield
an unexpected microscopic magnetic model for
volborthite, moving away from the kagome model. Moreover, we
reveal strong similarities to the physics of frustrated
coupled chains due to OO. Our subsequent simulations of the
microscopic model evidence an improved agreement with the
experimental data.

\section{DFT calculations}

The DFT calculations have been performed in the local
density approximation (LDA) using the full
potential\cite{footnote8} code
\texttt{fplo8.65-32}.\cite{fplo} For the scalar
relativistic calculations, the Perdew and Wang
parametrization\cite{PW} of the exchange-correlation
potential has been used. All calculations have been
performed on well-converged $k$-meshes.\cite{footnote9}

The reliability of DFT calculations depends crucially on the
accuracy of the experimental structural data used as input.
The chemical composition of volborthite hampers structural
studies due to the considerable content of V and H atoms,
which are poor scatterers of neutrons and x-rays,
respectively. Therefore, prior to investigations of subtle
electronic effects, the structural data should be addressed.
Among several structural data sets available we have chosen
a structural model (exp) based on joint x-ray and neutron
diffraction studies.  Although such combination improves the
reliability of the resulting data, the statistics (number of
reflections) is not sufficient for fully conclusive results.
Therefore, we have carried out a structural optimization
relaxing the atomic coordinates and minimizing the forces,
since LDA calculations for cuprates usually yield accurate
and consistent
results.\cite{kapel_hayd_PRL,sqlat_anisotr_model,HC_CuSe2O5_DFT_chiT_CpT_simul} Moreover,
to evaluate the influence of the different structural
models, we perform calculations for the experimental as well
as for the optimized structure.

\begin{figure}[tb]
\includegraphics[width=8.6cm]{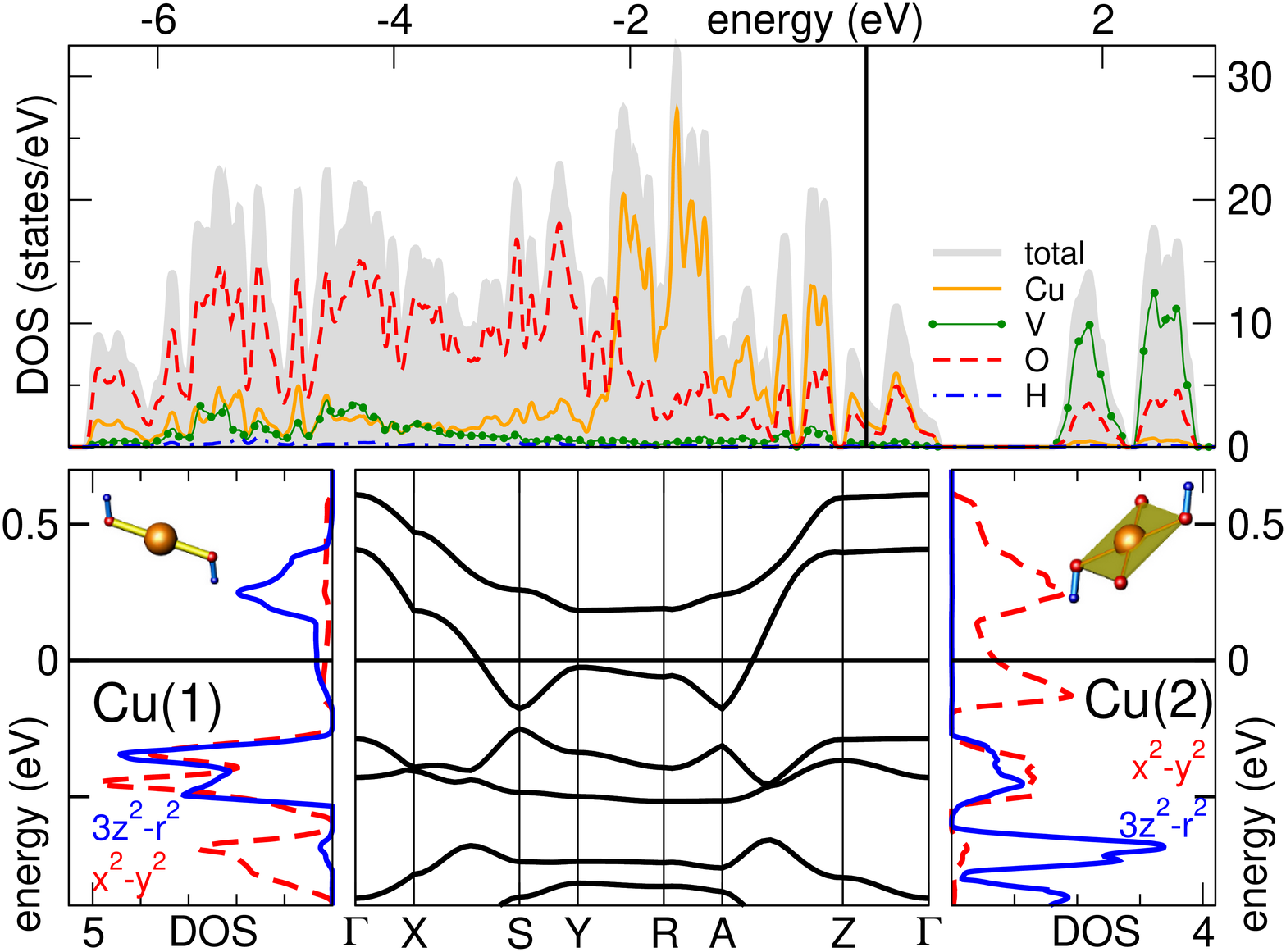}
\caption{\label{dosband}(Color online) Density of states
(upper panel), the band structure (lower panel, center) and
the orbital-resolved density of states for Cu(1) (left) and
Cu(2) (right) of the optimized structure ($3d_{x^2-y^2}$ is shown with a
dashed line).}
\end{figure}

The LDA-optimized crystal structure (opt) yields considerably
lower total energy and atomic forces. Although individual
bond lengths and angles change up to several percents (O--H
distance increased by $\sim$10\%), the overall structural
motive and the different local environment of Cu(1) and
Cu(2) are inherited from the original model.  

LDA yields a valence band width of 7 eV (Fig.~\ref{dosband})
typical for cuprates, and a metallic GS in contrast to the
green, transparent samples. This well-known problem of the
LDA originates from the underestimation of strong on-site
correlations for a Cu $3d^9$ configuration. Nevertheless,
LDA is a reliable tool to evaluate the relevant orbitals and
couplings.\cite{footnote9} For most cuprates, an effective one-band model
is well justified by a band complex at Fermi level
($\varepsilon_{\mathsf{F}}$) formed by $N$ antibonding
bands, where $N$ is the number of Cu atoms per cell.
In
contrast, in volborthite ($N$~=~3), six bands in vicinity of
$\varepsilon_{\mathsf{F}}$ (Fig.~\ref{dosband}) evidence
a sizable hybridization of two different $3d$ orbitals at each Cu
site that need to be included into the modeling.

The relevant Cu $3d$ orbitals are revealed by projecting the
density of states (DOS) onto local orbitals.  The resulting
orbital-resolved DOS is shown in Fig.~\ref{dosband}. For
both Cu(1) and Cu(2), the $3d_{x^2-y^2}$ and $3d_{3z^2-r^2}$
states are relevant and hybridized with each other. To
evaluate the relevant couplings, we consider two orbitals
($3d_{x^2-y^2}$ and $3d_{3z^2-r^2}$) per Cu atom and fit the
six bands using the Wannier functions (WFs)
technique.\cite{WF_ref,footnote9}

Prior to evaluation of the relevant couplings, the correct
orbital GS should be found. LDA yields an essentially
different filling of the orbitals: for Cu(1) the
$3d_{3z^2-r^2}$ is close to half-filling and the
$3d_{x^2-y^2}$ is almost filled (Fig.~\ref{dosband}, left),
while for Cu(2) it is the other way round
(Fig.~\ref{dosband}, right). However, the closer proximity
to half-filling in the LDA picture does not necessarily
provide the correct answer, as revealed for the related
system CuSb$_2$O$_6$.\cite{CuSb2O6_DFT}  Thus, we
cross-check the LDA result by LSDA+$U$ calculations.  In
agreement with the LDA, the latter yield the magnetically
active Cu(1) $3d_{3z^2-r^2}$ and Cu(2) $3d_{x^2-y^2}$ (see
details below).

The relevant transfer integrals can be extracted from the WFs
considering the hoppings between the GS orbitals (Cu(1)
$3d_{3z^2-r^2}$ and Cu(2) $3d_{x^2-y^2}$). The leading terms
are $t_1$ and $t_{\mathsf{ic}}$, which coincide with two NN
couplings in the AKM (Fig.~\ref{str}, bottom).  Surprisingly,
the next-nearest-neighbor (NNN) coupling $t_2$ is also
sizable, while other couplings are considerably smaller.
A small value of the inter-layer coupling supports the 2D
nature of magnetism.

The correct description of the orbital GS requires an
appropriate description of correlations in the Cu $3d$
shell, which can be treated in a mean-field way using the
LSDA+$U$ scheme. By stabilizing solutions comprising
different orbital occupations and a subsequent comparison of
their total energies, we evaluate the orbital GS. The
separation between the orbital GS and the lowest lying
excited orbital state ($3d_{x^2-y^2}$ for Cu(1) and Cu(2))
exceeds 500~meV ($\sim$6000~K), almost two orders of magnitude larger than
the magnetic exchange ($\sim$100~K). Therefore, orbital and spin degrees
of freedom are mostly decoupled and can be analyzed
separately.

\begin{figure}[tb]
\includegraphics[width=8.6cm]{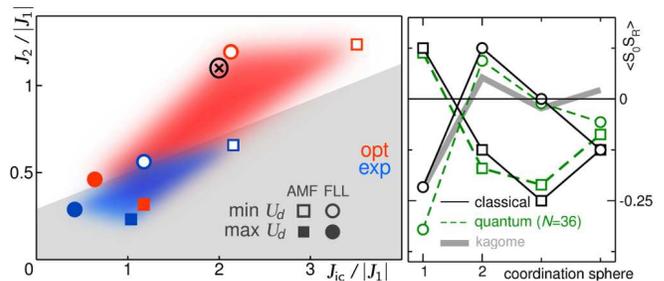}
\caption{\label{phase_diag}(Color online) Left: exchange
integrals as a function of the structural model (opt or exp), the LSDA+$U$
DCC scheme (AMF or FLL) and $U_d$
(min $U_d$, max $U_d$) on the phase diagram of the
$J_1$-$J_2$-$J_{\mathsf{ic}}$ model. The
white and gray fields
correspond to the singlet and the ferrimagnetic phases,
respectively. The shaded areas depict possible values of
exchange couplings for the both structural models. Our ED
results yield the best agreement for
$J_{\mathsf{ic}}$/$|J_1|$~=~2, $J_2$/$|J_1|$~=~1.1,
depicted by a cross. Right: correlation functions along
(squares) and between (circles) the $J_1$-$J_2$ chains
deviate significantly from the kagome model (bold gray line).}
\end{figure}

The leading exchange integrals $J_1$, $J_2$ and
$J_{\mathsf{ic}}$ are obtained mapping the results of
LSDA+$U$ total energy calculations onto a Heisenberg
model.\cite{footnote2} A careful analysis of the results
shows that the individual values of exchange integrals are
sensitive to (i) the structural model, (ii) the Coulomb
repulsion $U_d$ and  (iii) the double-counting correction
(DCC) scheme.\cite{footnote3}  The crucial influence of
these parameters is visualized in Fig.~\ref{phase_diag},
where the results for the experimental and optimized
structures are shown. For each structural model, we use the
limiting cases for the DCC --- around-the-mean-field (AMF) and
the fully localized limit (FLL)\cite{LDA_U_AMF_FLL} and vary
$U_d$ within a reasonable range.\cite{footnote4} Depending
on these parameters, we obtain $J_1$~=~$-$80$\pm10$~K,
$J_2$~=~35$\pm15$~K  and $J_{\mathsf{ic}}$~=~100$\pm60$~K
for the experimental and $J_1$~=~$-$65$\pm15$~K,
$J_2$~=~45$\pm15$~K and $J_{\mathsf{ic}}$~=~100$\pm60$~K for
the optimized lattice. 

The substantially ferromagnetic nature of $J_1$, in accord with
Goodenough-Kanamori-Anderson rules, and the relatively small
uncertainties of its strength disregarding the parameters
used give strong evidence that the pure kagome model is
inappropriate for volborthite.  The AKM can be ruled out
since $J_1$ and $J_{\mathsf{ic}}$ support each other and do
not give rise to frustration. Our microscopic insight
evidences that an essentially different model with
frustration governed by NNN exchange $J_2$ (competing with
both $J_1$ and $J_{\mathsf{ic}}$) should be used for
volborthite.  Despite sizable scattering of the $J$ values,
important general trends can be established. First, the
optimized structure has an enhanced $J_2$/$|J_1|$ ratio
compared to the experimental structure.  Second, FLL yields
considerably smaller $J_{\mathsf{ic}}$ and somewhat larger
values for $J_2$ than AMF.

Based on DFT calculations, we obtain a microscopic magnetic
model and determine the parameters $J_1$, $J_2$ and
$J_{\mathsf{ic}}$. Although we find the relevant region in
the phase space, the complexity of volborthite impedes a
more accurate estimate of individual exchange integrals,
especially $J_{\mathsf{ic}}$. In this case, refining the
parameters by numerical simulations of measured physical
properties and subsequent comparison to experimental data is
an appropriate way towards a deeper understanding.

\section{Exact diagonalization}

To realize a guided search for a consistent set of exchange
integrals, the GS of the $J_1$-$J_2$-$J_{\mathsf{ic}}$ model
should be investigated.  We explore the phase space by
considering a Heisenberg model with $J_1<0$, $J_2>0$,
$J_{\mathsf{ic}}>0$.  On a classical level, we find two GS:
a ferrimagnetic (fM) phase with magnetization $m=1/3$ and an
incommensurate $m=0$ helical (H) phase with spiral
correlations along $J_1$-$J_2$ frustrated chains, similar to
those of edge shared quasi 1D cuprates (see e.g.
Ref.~\onlinecite{Li2ZrCuO4_chiT_CpT_DFT_simul}).  The
transition from the fM-phase to the H-phase is driven by the
frustrating NNN in-chain coupling $J_2$ and occurs at
$J^{class}_2=|J_1|/4+J_{ic}/8$.  To discuss the GS of the
quantum model we use Lanczos exact diagonalization of finite
lattices up to $N=36$ sites.\cite{jr1,jr2}  For the quantum model, the fM
state competes with a singlet GS with $m=0$, and the
transition is given by
$J^{quant}_2=0.304|J_1|+0.200J_{ic}$.\cite{footnote5} The
transition line together with the DFT-derived exchange
integrals are plotted in Fig.~\ref{phase_diag} (left). Since
the experiments evidence zero magnetization of the GS, the
fM solutions can be ruled out and the analysis can be
restricted to the singlet GS. 

\begin{figure}[tb]
\includegraphics[angle=270,width=7.0cm]{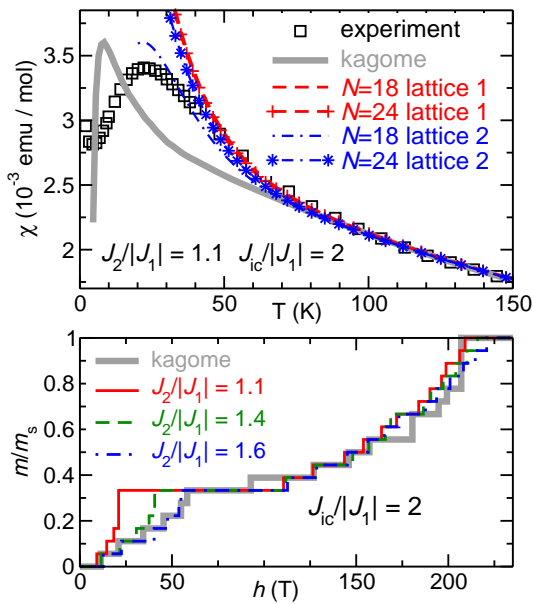}
\caption{\label{m_h}(Color online) Top: fits to the experimental
$\chi(T)$.\cite{footnote1} The solution of the $J_1$-$J_2$-$J_{\mathsf{ic}}$
model yields an improved description down to 50~K compared to the kagome model
(bold gray line).  Bottom: magnetization curves ($N$=36 sites) for different
solutions of the $J_1$-$J_2$-$J_{\mathsf{ic}}$ model in comparison to the
kagome model.}
\end{figure}

To understand the nature of the GS, we consider spin
correlations as a sensitive probe for magnetic ordering.
While the correlations between the chains are similar to the
standard kagome Heisenberg antiferromagnet, they are
completely different along the $J_1$-$J_2$ frustrated chains
(Fig.~\ref{phase_diag}, right).  These in-chain correlations
fit to a spiral state with a pitch angle very close to
the classical model. Hence, our data suggest well
pronounced in-chain spiral correlations together with weaker
inter-chain correlations. We mention, however, that these
statements are restricted to short-range correlations.

Since the magnetic correlations along the chains are
strongest, one could argue that the model exhibits an
effectively 1D low-temperature physics as has been discussed
previously for other 2D models such as the crossed-chain
model,\cite{Starykh_PRL_02} anisotropic triangular
lattice,\cite{Coldea_PRL_01} as well as for modified kagome
compounds.\cite{kapel_hayd_PRL} However, this issue as well
as a conclusive answer to the question of helical LRO need
further investigation.

To comprise the experimental magnetization curve, we add the
magnetic field term to the Heisenberg Hamiltonian and
simulate the $m(h)$ dependence. For the boundary of the fM
and singlet GS, we find a wide 1/3 magnetization plateau
starting at $h_{c1}=0$. However, the modification of the
exchange parameters, in particular increasing of $J_2$ and
decreasing of $|J_1|$ and $J_{\mathsf{ic}}$, according to the limits
set by DFT calculations leads to a significant increase of
$h_{c1}$ and to a drastic diminishing of the plateau width.
Close to the DFT-boundary ($J_{\mathsf{ic}}$/$|J_1|$~=~2,
$J_2$/$|J_1|$~=~1.1, $J_{\mathsf{ic}}$~=~100~K), we obtain
$h_{c1}$~=~22~T, which is still smaller than the
experimentally observed value.  We should note that this
deviation originates from the minimalistic character of the
model and considerable finite size effects.  Nevertheless, a
slightly modified ratio $J_2$/$|J_1|$~=~1.6 yields
$h_{c1}$~=~55~T (Fig.~\ref{m_h}, bottom) in excellent agreement with
the experiment. It should be mentioned that the nature of
spin correlations in the 1/3-plateau phase is substantially
different compared to the kagome
model.\cite{kagome_GS_field_plateau} Small magnetization
jumps seen experimentally\cite{volb_magn_high_field} can
not be resolved with present lattice sizes and might be
related to anisotropic exchange.  

We also calculate the temperature dependence of magnetic susceptibility
$\chi(T)$ using two different lattices\cite{footnote9} up to $N=24$ sites.  We
obtain a good fit down to 50~K (Fig.~\ref{m_h}, top),\cite{footnote7} the
resulting $g$~=~2.16 and $J_{\mathsf{ic}}$~=~100.5~K are in excellent agreement
with experiments\cite{volb_high_field_ESR} and our estimates from DFT.

\section{Summary}
To summarize, we suggest a new magnetic model for
volborthite: assuming that DFT calculations are applicable
to volborthite as they are for a plethora of compounds, the
kagome model can be safely ruled out. Instead, the magnetism
of volborthite is accounted for by a
$J_1$-$J_2$-$J_{\mathsf{ic}}$ model reminiscent of coupled
frustrated chains. For the proposed model, the orbital order
of Cu(1) $3d_{3z^2-r^2}$ and Cu(2) $3d_{x^2-y^2}$ orbitals
is crucial.

We suggest new experiments to challenge our model: resonant x-ray scattering measurements
measurements to study orbital effects and measurements in
high magnetic fields ($>$70~Tesla) to get an access to the
magnetization plateau.  Additional investigations of the
$J_1$-$J_2$-$J_{\mathsf{ic}}$ model itself by alternative
simulation methods are desirable to clarify the influence of
finite size effects, intrinsic for exact diagonalization.

\section*{ACKNOWLEDGEMENTS}
We thank Z.~Hiroi for providing us with $\chi(T)$ data. We
acknowledge fruitful discussions and valuable comments from
Z.~Hiroi, F.~Mila, G.~Nilsen and A.~Tsirlin.

\newpage
\renewcommand{\thefigure}{S\arabic{figure}}
\renewcommand{\thetable}{S\arabic{table}}
\setcounter{figure}{0}
\setcounter{table}{0}
\begin{widetext}
\begin{center}
\centerline{\textbf{Supplementary material}}
\end{center}

\begin{figure}[h]
\includegraphics[width=10.0cm]{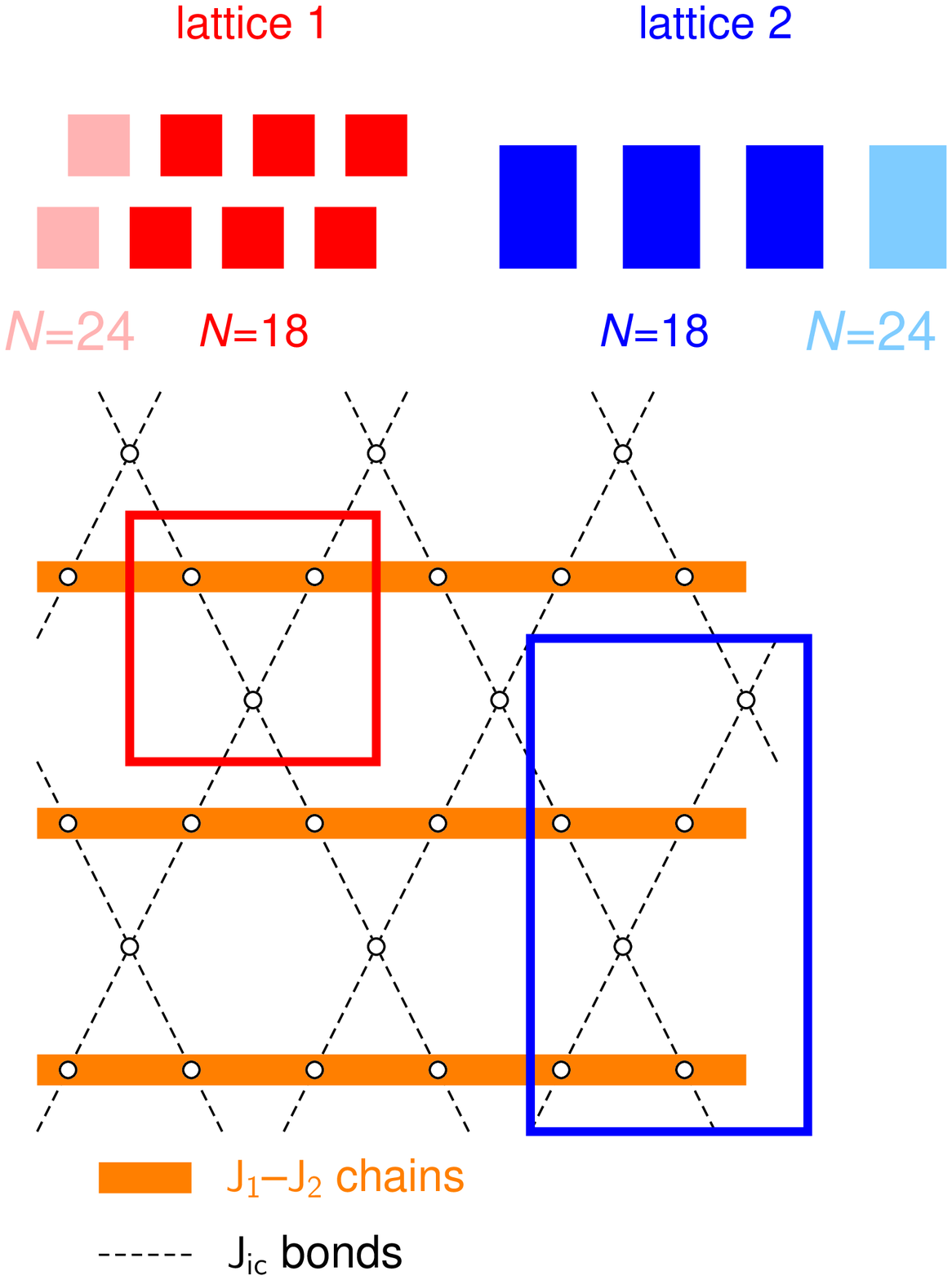}
\caption{\label{FL}Finite lattices used for exact diagonalization on $N=18$
and $N=24$ sites.}
\end{figure}
\smallskip

\begin{table}[h]
\caption{\label{basis_table}Basis set used in fplo8.65-32.}
\begin{ruledtabular}
\begin{tabular}{c c c c c}
atom & \multicolumn{4}{c}{states} \\
& core & valence & second valence & polarization \\
Cu & $1s$ $2s$ $2p$ & $3s$ $3p$ $4s$ $3d$ & $5s$ $4d$ & $4p$ \\
V & $1s$ $2s$ $2p$ & $3s$ $3p$ $4s$ $3d$ & $5s$ $4d$ & $4p$ \\
O &  $1s$ & $2s$ $2p$ & $3s$ $3p$ & $3d$ \\
H &  & $1s$ & $2s$ & $2p$ \\
\end{tabular}
\end{ruledtabular}
\end{table}
\smallskip

\begin{figure}[h]
\includegraphics[angle=270,width=10.0cm]{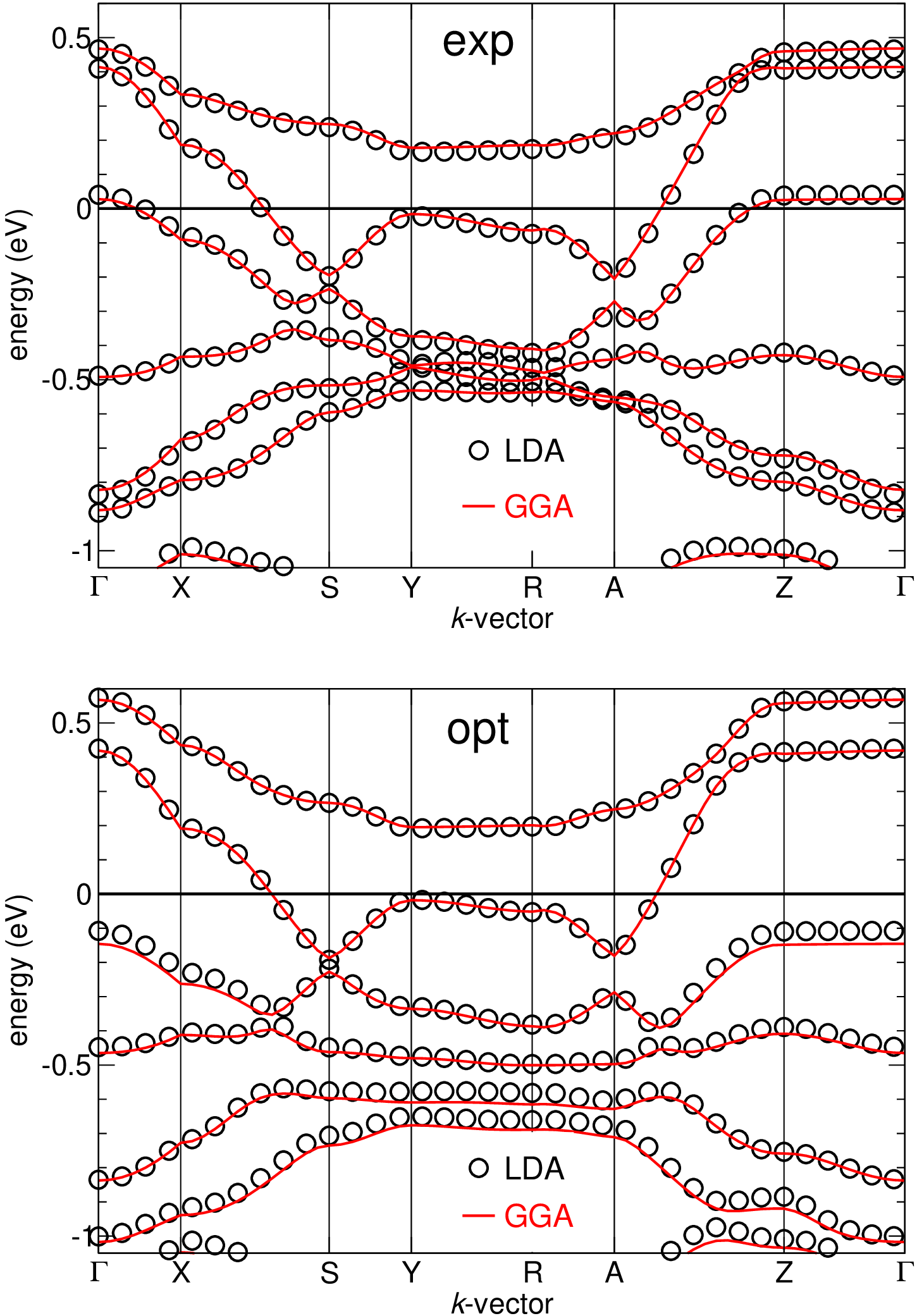}
\caption{\label{BS_LDA_GGA}Comparison of the band structures calculated within the local density approximation (LDA) and the general gradient approximation (GGA) for the experimental (``exp'', top panel) and optimized (``opt'', bottom panel) structures.}
\end{figure}
\medskip

\begin{figure}[h]
\includegraphics[angle=270,width=10.0cm]{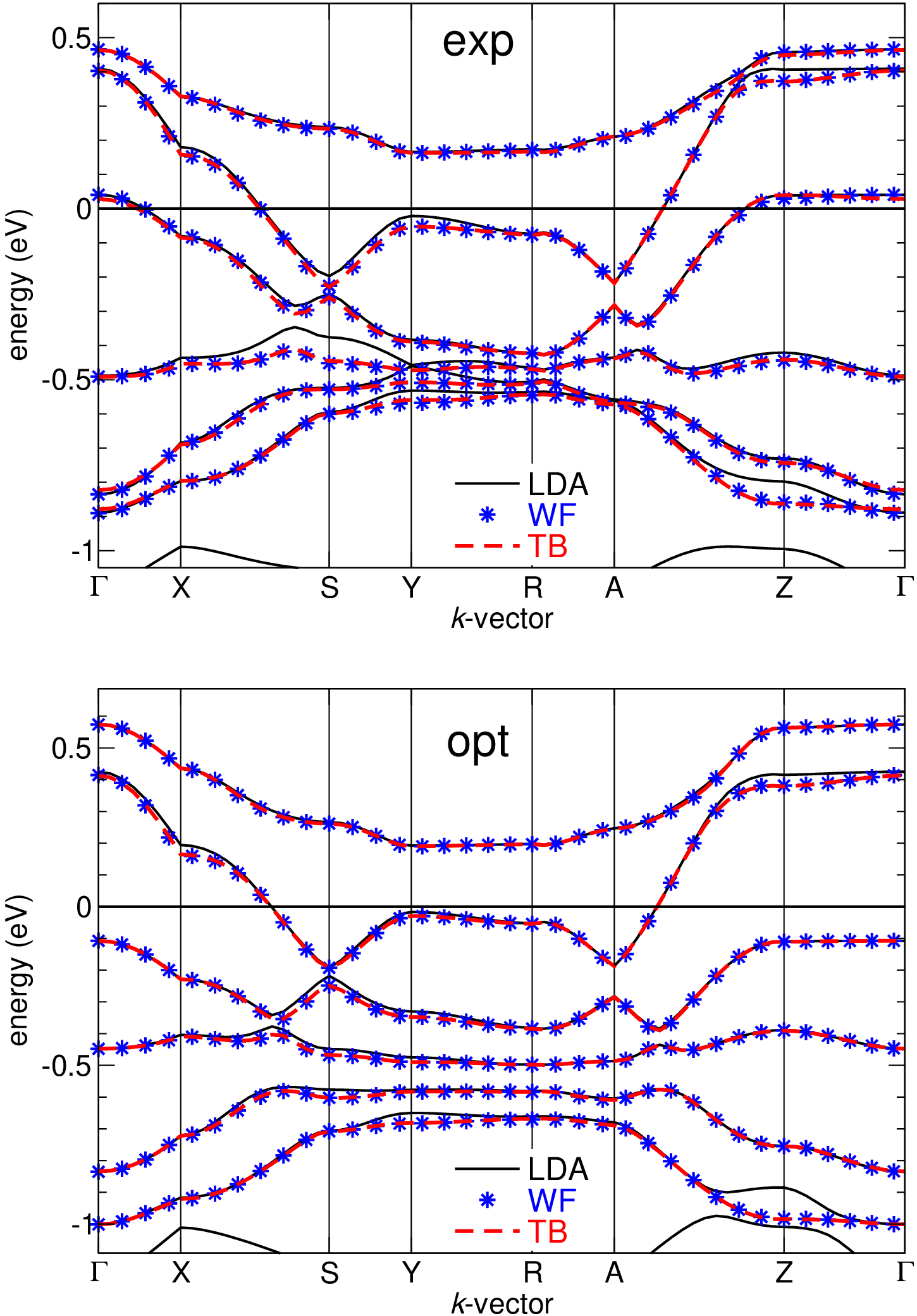}
\caption{\label{LDA_WF_TB}The Wannier functions fit (WF) and the tight-binding fit (TB) together with the LDA band structure for the experimental (``exp'', top panel) and optimized (``opt'', bottom panel) structures.}
\end{figure}
\medskip

\begin{table}[h]
\caption{\label{cells_meshes_table}Cells used for LDA and GGA (DFT) as well as spin-polarized
LSDA+$U$  and GGA+$U$ (DFT+$U$) calculations. For each cell, the total number (tot.) as well as the number of symmetrically inequivalent (ineq.)  Cu atoms are provided. Basis vectors are given in terms of the unit
cell vectors $\vec{a}$, $\vec{b}$ and $\vec{c}$. PUC --- primitive unit cell, SC --- supercell.}
\begin{ruledtabular}
\begin{tabular}{c c c c c c c c c}
cell & Sp. gr. & functional &  \multicolumn{2}{c}{Cu atoms} &  \multicolumn{3}{c}{basis vectors} & $k$-mesh\\
& & & tot. & ineq. & & & & \\
PUC & $P2/m$ & DFT & 3 & 2 & $0.5(\vec{a}+\vec{b})$ & $0.5(\vec{a}-\vec{b})$ & $\vec{c}$ & 12$\times$12$\times$12 \\
SC1 & $P\bar{1}$ & DFT & 3 & 3 & $0.5(\vec{a}+\vec{b})$ & $0.5(\vec{a}-\vec{b})$ & $\vec{c}$ & 8$\times$8$\times$8 \\
    &  & DFT+$U$ & & & & & & 4$\times$4$\times$4 \\
SC2 & $P\bar{1}$ & DFT & 6 & 6 & $0.5(\vec{a}+\vec{b})$ & $(\vec{a}-\vec{b})$ & $\vec{c}$ & 6$\times$3$\times$6 \\
    &  & DFT+$U$ & & & & & & 3$\times$2$\times$3 \\
SC3 & $P\bar{1}$ & DFT & 6 & 6& $\vec{a}$ & $\vec{b}$ & $\vec{c}$ & 4$\times$7$\times$5 \\
    &  & DFT+$U$ & & & & & & 2$\times$4$\times$3 \\
\end{tabular}
\end{ruledtabular}
\end{table}

\begin{table}[h]
\caption{\label{t_table}Transfer integrals $t$ as a function of the structural model (exp or opt) and the exchange-correlation potential (LDA or GGA), calculated using the Wannier functions technique. The numbers given correspond to hoppings between the magnetically active orbitals (Cu1 $3d_{3z^2-r^2}$ and Cu2 $3d_{x^2-y^2}$).}
\begin{ruledtabular}
\begin{tabular}{c c c c c}
 & \multicolumn{2}{c}{exp} & \multicolumn{2}{c}{opt} \\
 & LDA & GGA & LDA & GGA \\
$t_1$, meV & 91 & 93 & 117 & 119 \\
$t_2$, meV & 59 & 57 & 64 & 62 \\
$t_{\text{ic}}$, meV & 156 & 156 & 155 & 157 \\
\end{tabular}
\end{ruledtabular}
\end{table}

\end{widetext}  

\begin{thebibliography}{33}
\bibitem[{\citenamefont{footnote 2}()}]{footnote2}
\bibfnamefont{Only spin degrees of freedom are varied,
keeping the orbital GS preserved}.
\bibitem[{\citenamefont{footnote 3}()}]{footnote3}
\bibfnamefont{The situation is even more complicated because
the choice of $U_d$ depends on the relevant
orbital~\cite{CuSb2O6_DFT} and thus different values of
$U_d$ should be used for Cu(1) and Cu(2)}.
\bibitem[{\citenamefont{footnote 4}()}]{footnote4}
\bibfnamefont{According to extensive studies
of related Cu$^{2+}$ materials, ${U_d}$=10$\pm$1.5~eV for Cu(1) and
${U_d}$=7$\pm$1.5~eV for Cu(2) were chosen as typical values
with sizable error bars for the resulting
exchange integrals to account for the uncertainties in
${U_d}$. An extensive study of the details including the DCC
will be published elsewhere}.
\bibitem[{\citenamefont{footnote 5}()}]{footnote5}
\bibfnamefont{Note that due
to boundary effects also for the classical model on that
finite lattice the expression given above is modified to
$J^{class}_2=0.340|J_1|+0.166J_{ic}$}.
\bibitem[{\citenamefont{footnote 7}()}]{footnote7}
\bibfnamefont{At low temperatures, finite size effects are strong as evidenced
by a sizable deviation between the results for different lattices as well as
between $N=18$ and $N=24$ curves for the lattice 2}.
\end{thebibliography}
\end{document}